\documentclass[12pt]{article}
\usepackage{cite}
\input epsf

\newcommand{\sect}[1]{\setcounter{equation}{0}\section{#1}}
\newcommand{\scri}{I}

\begin{document}

\title{Holography in General Space-times}

\author{{\sc Raphael Bousso}\thanks{\it
    bousso@stanford.edu} \\[.3 ex]
    {\it Department of Physics, Stanford University,}\\
    {\it Stanford, California 94305-4060}
%\\[1.4ex]
%{\sc ANOTHER AUTHOR}\thanks{\it aaa@bbb.edu}
%\\[.3 ex] {\it AAA Institute, University of BBB},\\
%  {\it CITY, XX 12345-6789}
%
}

\date{SU-ITP-99-24~~~~~3 June 1999~~~~~hep-th/9906022}

\maketitle

\begin{abstract}

We provide a background-independent formulation of the holographic
principle.  It permits the construction of embedded hypersurfaces
(screens) on which the entire bulk information can be stored at a
density of no more than one bit per Planck area.  Screens are
constructed explicitly for AdS, Minkowski, and de~Sitter spaces with
and without black holes, and for cosmological solutions.  The
properties of screens provide clues about the character of a
manifestly holographic theory.

\end{abstract}

\pagebreak

%%%%%%%%%%%%%%%%%%%%%%%%%%%%%%%%%%%%%%%%%%%%%%%%%%%%%%%%%%%%%%%%%%%%
\sect{Holographic Principle}
\label{sec-intro}
%%%%%%%%%%%%%%%%%%%%%%%%%%%%%%%%%%%%%%%%%%%%%%%%%%%%%%%%%%%%%%%%%%%%

\subsection{The holographic principle for general spacetimes}
\label{sec-hp}

In Ref.~\cite{Bou99b} a covariant entropy bound was conjectured to
hold in general space-times.  The bound can be saturated, but not
exceeded, in cosmology and in collapsing regions.  Applied to finite
systems of limited self-gravity, it reduces to Bekenstein's
bound~\cite{Bek81}.  For a $D$-dimensional Lorentzian space-time, the
covariant bound can be stated as follows:

\paragraph{Covariant Entropy Conjecture}
{\em Let $A$ be the area of a connected $(D-2)$-dimensional spatial
surface $B$.  Let $L$ be a hypersurface bounded by $B$ and generated
by one of the four null congruences orthogonal to $B$.  Let $S$ be the
total entropy contained on $L$.  If the expansion of the congruence is
non-positive (measured in the direction away from $B$) at every point
on $L$, then $S \leq A/4$.}\\[1.5ex]
The conjecture can be viewed as a generalization of an entropy bound
proposed by Fischler and Susskind~\cite{FisSus98}.  It differs in that
it considers all four light-like directions and selects some of them
by the criterion of non-positive expansion.  Several concepts crucial
to a light-like formulation were recognized earlier by Corley and
Jacobson~\cite{CorJac96}, who distinguished between ``past and future
screen maps'' and drew attention to the importance of caustics.  We
should also point out a number of recent proposals for entropy bounds
in cosmology~\cite{EasLow99,Ven99,BakRey99,KalLin99,Bru99} (see
Sec.~\ref{sec-frw}).

The covariant entropy bound is manifestly invariant under time
reversal.  This property cannot be understood if the bound applies
only to thermodynamic entropy.  One is thus forced to interpret the
bound as a limit on the number of degrees of freedom ($N_{\rm dof}$)
that constitute the statistical origin of any thermodynamic entropy
that may be present on $L$.  Since no assumptions about the
microscopic properties of matter were made, the limit is
fundamental~\cite{Bou99b}.  There simply cannot be more independent
degrees of freedom on $L$ than $A/4$, in Planck units.  This
conclusion compels us to embrace the holographic conjecture of
't~Hooft~\cite{Tho93} and Susskind~\cite{Sus95}, and it motivates the
following background-independent formulation of their hypothesis:

\paragraph{Holographic Principle}
{\em Let $A$ be the area of a connected $(D-2)$-di\-men\-sion\-al
spatial surface $B$.  Let $L$ be a hypersurface bounded by $B$ and
generated by one of the four null congruences orthogonal to $B$.  Let
$\cal N$ be the number of elements of an orthonormal basis of the
quantum Hilbert space that fully describes all physics on $L$.  If the
expansion of the congruence is non-positive (measured in the direction
away from $B$) at every point on $L$, then ${\cal N} \leq
e^{A/4}$.}\\[1.5ex]%
Simplifying slightly,%
\footnote{We thank Gerard 't~Hooft for suggesting a formulation in
terms of Hilbert space.}
one could state that {\em $N_{\rm dof} \leq A/4$, where $N_{\rm dof}$
is the total number of independent quantum degrees of freedom present
on $L$.}\\[0.5ex]

The holographic principle thus assigns at least two light-like
hypersurfaces to any given spatial surface $B$ and bounds $N_{\rm
dof}$ on those hypersurfaces.  The relevant hypersurfaces can be
constructed as follows (see Ref.~\cite{Bou99b} for a more detailed
discussion).  There will be four families of light-rays orthogonal to
$B$: a past-directed and a future-directed family on each side of $B$.
Consider only the families with non-positive expansion away from $B$.
Generically there will be two such families, but if the expansion is
zero in some directions, there may be as many as three or four.  Pick
one of the allowed families and follow each light-ray until the
expansion becomes positive or a boundary of space-time is reached.
The null hypersurface thus generated is called a {\em light-sheet}.
$N_{\rm dof}$ on a light-sheet of $B$ will not exceed a quarter of the
area of $B$.

In general, the holographic principle associates the area of a surface
$B$ with $N_{\rm dof}$ on null hypersurfaces, not spatial regions,
bounded by $B$.  Under certain conditions, however, the bound does
apply to space-like hypersurfaces as well.  This was shown in
Ref.~\cite{Bou99b} for the covariant entropy bound.  For the
holographic principle, the derivation can be repeated, with
``entropy'' replaced by ``$N_{\rm dof}$.''  It yields the following
theorem:

\paragraph{Spacelike Projection Theorem}
{\em Let $A$ be the area of a closed surface $B$ possessing a
future-directed light-sheet $L$ with no boundary other than $B$.  Let
the spatial region $V$ be contained in the intersection of the causal
past of $L$ with any spacelike hypersurface containing $B$.  Let
$N_{\rm dof}$ be the total number of independent quantum degrees of
freedom present on $V$.  Then $N_{\rm dof} \leq A/4$.}\\[1.5ex]%
The theorem can be widely applied and easily understood.  Under the
stated conditions, none of the degrees of freedom on $V$ can escape
through holes in $L$ or be destroyed on a singularity.  Then causality
and the second law of thermodynamics require all degrees of freedom on
$V$ to be present on $L$ as well.  On $L$ their number is bounded by
the holographic principle, so it must be bounded also on $V$.
%One
%can repeat this argument, replacing ``future'' everywhere by ``past''
%and vice versa, to derive a time-reversed version of the theorem.
%(This would not be possible if we were referring to thermodynamic
%entropy~\cite{Bou99b}).
The spacelike projection theorem will be of use in a number of
spacetimes, including de~Sitter and AdS (Sec.~\ref{sec-examples}).

\subsection{Outline}
\label{sec-outline}

The holographic principle is a relation between space-time geometry
and the number of degrees of freedom.  It is not equivalent to the
statement that there exists a conventional theory without gravity,
living on the boundary of a space-time region, with one degree of
freedom per Planck area, by which all bulk phenomena including quantum
gravity can be described.  The holographic principle is clearly
necessary for the existence of such a theory, but as we will argue
below, it is not sufficient.

The holographic principle does imply, however, that all information
contained on $L$ can be stored on the surface $B$, at a density of no
more than one bit per Planck area. (We neglect factors of $\ln 2$.)
We shall work with this interpretation.  For a given space-time, we
ask whether the information in the interior can be completely
projected (in accordance with our formulation of the holographic
principle) onto suitable hypersurfaces which will be called screens.
We are led to a construction (Sec.~\ref{sec-projection}) under which
the space-time is sliced into null hypersurfaces $L$.  Each light-ray
on $L$ is followed in the direction of non-negative expansion until
the expansion becomes zero.  This yields a preferred location for a
screen encoding the information on $L$.  By repeating this procedure
for every slice $L$, one obtains one or more screen-hypersurfaces.
They will either be located on the boundary or will be embedded in the
interior of the spacetime.  We establish conditions under which
projection along spacelike directions is possible.

In Sec.~\ref{sec-examples} we apply the construction to examples of
space-times, including Anti-de~Sitter space, Minkowski space,
de~Sitter space, cosmological solutions, and black holes.  We find
that AdS has highly special properties under our construction, as it
admits spacelike projection onto timelike screens of constant area.
For all examples we find that the information in the entire spacetime
can be projected onto preferred screens.  For spaces with black holes,
inequivalent slicings lead to different screen structures.  We relate
this to the question of information loss.

We discuss the structure of holographic screens in
Sec.~\ref{sec-discussion}, and draw some conclusions.  In a number of
examples, the screen-hypersurfaces are spacelike.  In other examples
they are null or timelike, with the spatial area depending on time.
One would not expect such screens to admit a conventional Lorentzian
quantum field theory with one degree of freedom per Planck area,
because the number of degrees of freedom would have to be
time-dependent.

This suggests that a distinction should be made between a {\em dual
theory}, and the {\em holographic theory}
(Sec.~\ref{sec-screentheories}).  In both types, the holographic
principle would be manifest.  A dual theory would be characteristic of
a certain class of space-times.  It would be a conventional theory
without gravity, living on the geometric background defined by a
holographic screen of the space-time and containing one degree of
freedom per Planck area.  It would complement, or be equivalent to, a
quantum gravity theory living in the bulk, and could thus be used to
describe bulk physics.  The conformal field theory on the boundary of
AdS~\cite{Mal97,GubKle98,Wit98a,SusWit98} is an example of a dual
theory.
%[MENTION ALSO THE SYSTEMS OF D-BRANES USED TO DESCRIBE BLACK
%HOLES (STROMINGER-VAFA AND CALLAN-MALDACENA)?]

The existence of a dual theory in a given class of space-times will
depend on certain properties of the projection and the screen which we
aim to expose.  More generally, the structure of screens points to a
fundamental theory in which quantum degrees of freedom are a derived
concept, and their number can change.  The theory must give rise to
gravity by permitting the unique reconstruction of space-time geometry
from the effective number of degrees of freedom in such a way that the
holographic principle is manifestly satisfied (Sec.~\ref{sec-ge}).  We
call this the holographic theory.  Clearly it cannot be a conventional
quantum field theory living on a pre-defined geometric background.
Perhaps an indication of its character can be gained from certain
proposals of 't~Hooft~\cite{Tho93,Tho99}.

\paragraph{Notation and conventions}

We work with $D$-dimensional Lorentzian manifolds $M$.
The terms {\em light-like} and {\em null} are used interchangeably.
Any $(D-1)$-dimensional submanifold $H \subset M$ is called a
{\em hypersurface} of $M$~\cite{HawEll}.
If $D-2$ of its dimensions are everywhere spacelike and the
remaining dimension is everywhere timelike (null, spacelike), $H$ is
called a {\em timelike (null, spacelike) hypersurface}.
By a {\em surface} we always refer to a $(D-2)$-dimensional spacelike
submanifold $B \subset M$; by {\em area} we mean the proper volume of
a surface.
By a {\em light-ray} we do not mean an actual electromagnetic wave or
photon, but simply a null geodesic.
We use the terms {\em null congruence}, and {\em family of
light-rays}, to refer to a congruence of null
geodesics~\cite{HawEll,Wald}.
The term {\em light-sheet} is defined in Sec.~\ref{sec-hp}; the terms
{\em projection}, {\em screen}, and {\em screen-hypersurface} in
Sec.~\ref{sec-screens}; {\em dual theory} and {\em holographic theory}
in Sec.~\ref{sec-screentheories}.
We set $\hbar = c = G = k = 1$.

\pagebreak

%%%%%%%%%%%%%%%%%%%%%%%%%%%%%%%%%%%%%%%%%%%%%%%%%%%%%%%%%%%%%%%%%%%%
\sect{Holographic projection}
\label{sec-projection}
%%%%%%%%%%%%%%%%%%%%%%%%%%%%%%%%%%%%%%%%%%%%%%%%%%%%%%%%%%%%%%%%%%%%

\subsection{Screens}
\label{sec-screens}

The construction described in the previous section answers the
following question: Given a surface $B$ of area $A$, what is the
hypersurface $L$ on which $N_{\rm dof}$ is bounded by $A/4$?  We will
now consider space-times globally, and ask a different question: which
surfaces store the information contained in the entire space-time?  To
answer this question, the above prescription should be inverted.
Given a null hypersurface $L$, one should follow the geodesic
generators of $L$ in the direction of non-negative expansion.  One can
stop anytime, but one must stop when the expansion becomes negative.
This procedure will be called {\em projection}.  The
$(D-2)$-dimensional spatial surface $B$ spanned by the points where
the projection is terminated will be called a {\em screen} of the
projection.  If the expansion vanishes on every point of $B$, it will
be called a {\em preferred screen}.

Preferred screens are of particular interest for a simple reason.  The
expansion of the projection typically changes sign on a preferred
screen $B$.  Therefore $B$ will be a preferred screen for projections
coming from two directions, e.g., the past-directed outgoing and
future-directed ingoing directions.  It will thus be particularly
efficient in encoding global information.  (Actually, there may be a
deeper reason why preferred screens play a special role.  We suspect
that they are precisely the surfaces for which the holographic bound,
$N_{\rm dof} \leq A/4$, is saturated.  This is suggested by
considerations in Sec.~4.2 of Ref.~\cite{Bou99b}.  There it was found
that the covariant entropy bound can be saturated on the future
light-sheet of an apparent horizon, but not on the future light-sheets
of smaller spheres inside the apparent horizon.  Under the projection
that generates those light-sheets, the apparent horizon is a preferred
screen.  This argument should be viewed with caution, however, because
there might be independent, practical reasons why the thermodynamic
entropy cannot be made as large as $N_{\rm dof}$ for the smaller
spheres.)

By following all generators of the null hypersurface $L$ in a
non-contracting direction to a screen, we obtain a projection of all
information on the hypersurface onto one or more screens, which may be
embedded in the hypersurface, or may lie on its boundary.  The number
of screens can be minimized by using preferred screens whenever
possible.

\subsection{Screen-hypersurfaces}
\label{sec-sh}

In order to project the information in a space-time $M$, our strategy
will be to slice $M$ into a one-parameter family of null
hypersurfaces, $\{ L \}$.  This will be possible in all examples we
consider.  Usually the slicing is highly non-unique, but the
symmetries of most space-times of interest reduce the number of
inequivalent slicings considerably.  To each slice $L$, we apply the
projection rule.  This procedure yields a number of one-parameter
families of $(D-2)$-dimensional screens.  Each family forms a
$(D-1)$-dimensional screen-hypersurface embedded in $M$ or located on
the boundary of $M$.  (This sounds a lot more complicated than it
is---see the ``recipe'' in Sec.~\ref{sec-recipe} and the figures in
Sec.~\ref{sec-examples} below.)  The screen-hypersurfaces can be
time-like, null, or space-like; in Sec.~\ref{sec-examples} examples of
each type will be found.  In general, the causal character can change
from time-like to space-like within the screen-hypersurface.

Usually it will be clear whether we are talking about a screen (a
spatial surface), or a $(D-1)$-dimensional hypersurface formed by a
one-parameter family of screens.  Therefore we will often refer to a
screen-hypersurface loosely as a ``screen'' of $M$.  If the
hypersurface consists of preferred screens, we call it a preferred
screen-hypersurface, or loosely a preferred screen of $M$.  If the
expansions of both independent pairs of orthogonal families of
light-rays vanish on a screen, it will be preferred under all four
projections that end on it.  We will call such a screen, and
hypersurfaces formed by such screens, {\em optimal}.

So far we have discussed only {\em null projection}, i.e., projection
of information along null hypersurfaces.  It is sometimes possible to
project information along spacelike hypersurfaces.  Namely, {\em
spacelike projection} of the information in a spatial region $V$ onto
a screen $B$ is allowed if $V$ and $B$ satisfy the conditions set
forth in the ``spacelike projection theorem'' (Sec.~\ref{sec-hp}).
This will be significant in a number of space-times, in particular in
de~Sitter and Anti-de~Sitter space.

\subsection{The recipe}
\label{sec-recipe}

To construct screens, one must slice a space-time into null
hypersurfaces.  In view of the spherical symmetry of all metrics
considered below, it will be natural to slice them into a family $\{ L
\}$ of light-cones centered at $r=0$.%
\footnote{In Minkowski-space, we will also consider a family of
light-rays orthogonal to a flat $(D-2)$-plane.}
The family can be parametrized by time.  This will leave two
inequivalent null projections, namely along past or future-directed
light-cones.  Often the light-cones will be truncated by boundaries of
the space-time and will not include $r=0$, but this does not matter.
In the case of spherical symmetry, one thus obtains the following
recipe for the construction of screens:

\begin{enumerate}

\item{Draw a Penrose diagram.  Every point represents a
$(D-2)$-sphere.  Each diagonal line represents a light-cone.  The two
inequivalent null slicings can be represented by the ascending and
descending families of diagonal lines.}

\item{Pick one of the two families.  Now the question is in which
direction to project along the diagonal lines.}

\item{Identify the apparent horizons, i.e., hypersurfaces on which the
expansion of the past or future light-cones vanishes.  They will
divide the space-time into normal, trapped, and anti-trapped regions.
In each region, draw a wedge whose legs point in the direction of
negative expansion of the cones.}

\item{On a given diagonal line (i.e., light-cone), project each point
towards the tip of the local wedge, onto the nearest point (i.e.,
sphere) $B_i$ where the direction of the tip flips, or onto the
boundary of space-time as the case may be.}

\item{Repeat for every line in the family.  The surfaces $B_i$ will
form (preferred) screen-hypersurfaces $H_i$.}

\end{enumerate}

Below we will strive to make these steps explicit by including two or
three Penrose diagrams for most examples.  In the first diagram, the
apparent horizons will be identified and the wedges placed.  For each
inequivalent family of light-cones, we will then provide a diagram in
which the projection directions are indicated by thick arrows.  We
invite the reader to verify that these directions are uniquely
determined by the wedges.  Screens will be denoted by thick points,
preferred screen-hypersurfaces by thick lines.

\pagebreak

%%%%%%%%%%%%%%%%%%%%%%%%%%%%%%%%%%%%%%%%%%%%%%%%%%%%%%%%%%%%%%%%%%%%
\sect{Examples}
\label{sec-examples}
%%%%%%%%%%%%%%%%%%%%%%%%%%%%%%%%%%%%%%%%%%%%%%%%%%%%%%%%%%%%%%%%%%%%

\subsection{Anti-de~Sitter space}
\label{sec-ads}

Type IIB string theory on the background $\mbox{AdS}_5 \times
\mathbf{S}^5$, with $N$ units of flux on the $\mathbf{S}^5$, appears
to be dual to $(3+1)$-dimensional $U(N)$ supersymmetric Yang-Mills
theory with 16 real supercharges~\cite{Mal97}.  One can consider this
theory to live on the boundary of the AdS space.  The correspondence
between bulk and boundary~\cite{GubKle98,Wit98a} relates infrared
effects in the bulk to ultraviolet effects on the
boundary~\cite{GubKle96,GubKle97,Wit98a}.  This feature was exploited
by Susskind and Witten~\cite{SusWit98} to show that the boundary
theory has only one degree of freedom per Planck area, as required by
the holographic principle in the traditional, ``spacelike'' form in
which it has often been expressed.

We wish to understand some of these properties from the perspective of
the general formulation of the holographic principle given in
Sec.~\ref{sec-hp}.  From this point of view, the bulk information is
projected along null directions in general, and along spacelike
directions only if certain conditions are met.  We will verify that
these conditions are indeed satisfied in AdS.  Moreover, we will find
that the boundary at spatial infinity is a preferred (and optimal)
screen under our construction.  Finally, we will note that AdS admits
screen-hypersurfaces of constant spatial area that encode their
space-time interior.  The concurrence of these properties is special
to AdS (and to some unstable solutions identified in
Sec.~\ref{sec-esu}), and may be a necessary condition for the
existence of the kind of duality that has been found in this
space-time.

Anti-de~Sitter space can be scaled into the direct product of an
infinite time axis with a unit spatial ball~\cite{SusWit98}.  In this
form it has the metric
\begin{equation}
ds^2 = R^2 \left[ - \frac{1+r^2}{1-r^2} dt^2 + \frac{4}{(1-r^2)^2}
\left( dr^2 + r^2 d\Omega^2 \right) \right].
\label{eq-ads-metric}
\end{equation}
The constant scale factor $R$ is the radius of curvature.  The
spacelike hypersurfaces are open balls given by $t=\mbox{const}, 0
\leq r < 1$.  The boundary of space is a two-sphere residing at $r=1$.
The proper area of spheres diverges as $r \rightarrow 1$.

\begin{figure}[htb!]
  \hspace{.13\textwidth} \vbox{\epsfxsize=.74\textwidth
  \epsfbox{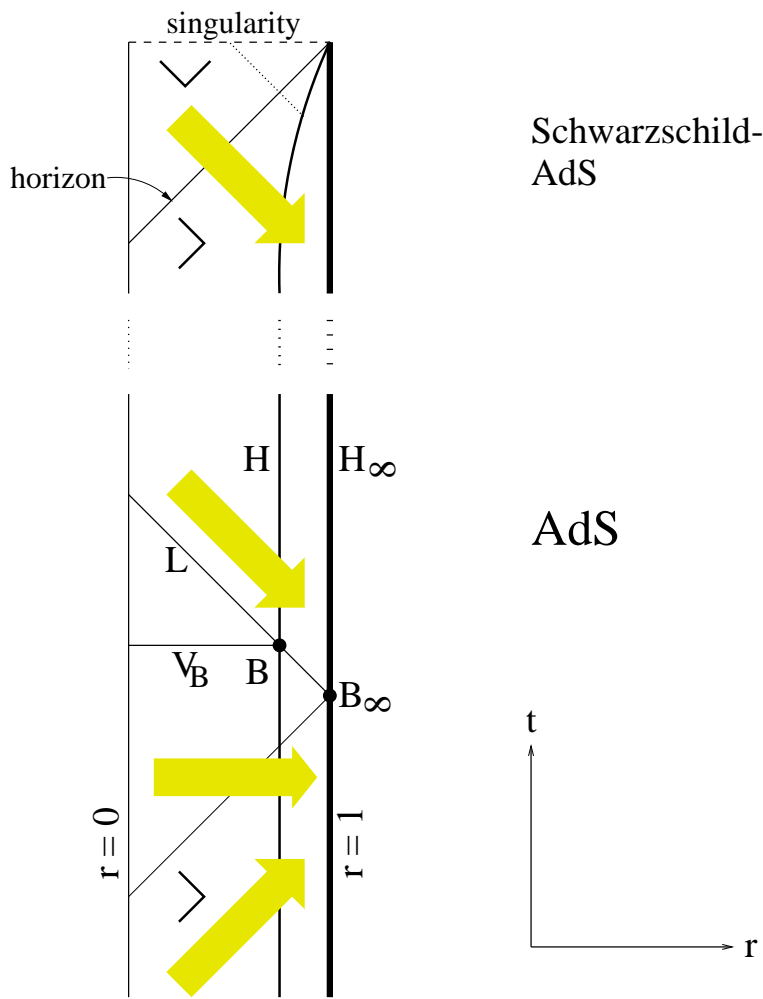}}
\caption%
{\small\sl Conventions and methods used in all diagrams are spelled
out in Sec.~\ref{sec-recipe}.  Anti-de~Sitter space contains no
apparent horizons; all spheres are normal.  Spacelike projection is
allowed.  All null and spacelike projections are directed away from
the center at $r=0$.  Interior information can thus be projected onto
a screen-hypersurfaces $H$ of constant area; $H$ encodes no exterior
information.  The screen at spatial infinity, $H_\infty$, is optimal
and encodes all bulk information. --- The upper part of the figure
shows a diagram for Schwarzschild-AdS.  Since the future light-sheets
of the screen surfaces are not complete (see dotted line), the black
hole interior cannot be projected onto $H$ along space-like
directions, but only along past light-cones.}
\label{fig-ads}
\end{figure}

Consider the past directed radial light-rays emanating from a caustic
($\theta = +\infty$) at $r=0, t=t_0$ (Fig.~\ref{fig-ads}).
They form a past light-cone $L$ with a spherical boundary.  The cone
grows with affine time until the light-rays reach the boundary of
space at $r=1$.  It is straightforward to check that the expansion,
$\theta$, is inversely proportional to the affine time.  It thus
decreases monotonically, but remains positive; one finds that
\begin{equation}
\theta \rightarrow 0~~~\mbox{as}~~~r\rightarrow 1.
\label{eq-theta-adsboundary}
\end{equation}

Consider a sphere $B$, of area $A_B$, on the lightcone $L$.  The part
of $L$ in the interior of $B$, $L_B$, has negative expansion in the
direction away from $B$, and therefore constitutes a light-sheet of
$B$.  By the holographic principle, the number of degrees of freedom
on $L_B$ does not exceed a quarter of the area of $B$:
\begin{equation}
N_{\rm dof}(L_B) \leq \frac{A_B}{4}.
\end{equation}

Because the cone closes off at $r=0$, it has no boundary other than
$B$.  The spatial interior of $B$ on any spacelike hypersurface
through $B$, $V_B$, lies entirely in the causal past of $L_B$.
Therefore the conditions for the spacelike projection theorem are met.
It follows that area bounds $N_{\rm dof}$ on spatial regions of AdS:
\begin{equation}
N_{\rm dof}(V_B) \leq \frac{A_B}{4}.
\end{equation}

Since the light-cone expansion is positive for all values of $r$,
these conclusions remain valid in the limit as the sphere $B$ moves to
the boundary of space, $B \rightarrow B_\infty$.  By
Eq.~(\ref{eq-theta-adsboundary}), $B_\infty$ is a preferred screen.
As expected, the preferred screen is precisely the one which encodes
the entire space.  By time reversal invariance of
Eq.~(\ref{eq-ads-metric}), the expansion of future-directed radial
lightrays arriving at $B_\infty$ will also vanish; thus $B_\infty$ is
an optimal screen.

So far we have considered screens bounding $N_{\rm dof}$ on a
particular light-cone or spatial hypersurface.  A screen-hypersurface
encoding the entire space-time is obtained by repeating the
construction for every single light-cone in the slicing of AdS.  By
the time-translation invariance of Eq.~(\ref{eq-ads-metric}), this
repetition is trivial.  The family of finite screens $B(t)$ of
constant area $A_B$ thus forms a timelike screen-hypersurface $H$ of
topology $\mathbf{R} \times \mathbf{S}^{D-2}$.  By the holographic
principle, $N_{\rm dof}$ in the enclosed space-time region does not
exceed $A_B/4$.  From the spacelike projection theorem it follows that
it does not matter whether one counts degrees of freedom on null or on
spacelike hypersurfaces intersecting $H$.  After taking the limit
$B(t) \rightarrow B_\infty(t)$, one finds that the timelike boundary
at $r=1$, $H_\infty$, is an optimal screen of Anti-de~Sitter space.
It encodes the entire information in the bulk, by spacelike or null
projection.

Let us briefly discuss what happens when a black hole forms. This is
shown in the upper part of Fig.~\ref{fig-ads}.  Let us assume that the
constant screen area, $A_B$, is so large that the black hole never
engulfs the screen-hypersurface $H$ formed by the screens $B(t)$.
Then the past-directed ingoing light-sheet of any $B(t)$ lies outside
the event horizon, and has no boundary other than $B(t)$.  By
arguments similar to those leading to the spacelike projection
theorem~\cite{Bou99b}, this implies that $N_{\rm dof}$ in the region
between the black hole and $H$ never exceeds $A_B/4$.  Generic
space-like hypersurfaces passing through the interior of the black
hole, however, are not contained in the causal past of any complete
future-directed light-sheet of $B$ (see dotted line in
Fig.~\ref{fig-ads}).  The spacelike projection theorem does not apply
to those regions, and therefore the spacelike projection of the
interior of the event horizon onto $H$ is not possible.  Of course,
the entire black hole interior can be encoded on $H$ by null
projection along past light-cones.  (Alternatively, it can be
projected onto the apparent horizon; we discuss this in more detail in
Sec.~\ref{sec-mink} for the case of Schwarzschild black holes.)  This
discussion remains valid in the limit as $B \rightarrow B_\infty$, and
thus applies to the boundary of Schwarzschild-AdS at spatial infinity.

\subsection{Minkowski space}
\label{sec-mink}

We now turn to space-times which are asymptotically flat.  The
discussion of finite bound systems in Minkowski space does not differ
much from the treatment in AdS.  The space-time region occupied by
them can be projected onto a screen-hypersurface of topology
$\mathbf{R} \times \mathbf{S}^{D-2}$, formed by a spherical screen
circumscribing the system.  As long as no black holes form, the
projection can be spacelike.  A spherical screen of finite size is not
preferred unless the interior is on the verge of gravitational
collapse (see Sec.~\ref{sec-esu}).

A bound system can also be projected along past-directed light-rays
onto a remote flat plane.  All families of null-geodesics orthogonal
to the plane have zero expansion; therefore the screen is optimal.  It
can encode bulk information on both sides.  This projection was
originally proposed by Susskind~\cite{Sus95} and was further
investigated in Ref.~\cite{CorJac96}.  If the system does not contain
black holes, it can be projected onto the plane along future-directed
light-rays as well.

\begin{figure}[htb!]
  \hspace{.15\textwidth} \vbox{\epsfxsize=.7\textwidth
  \epsfbox{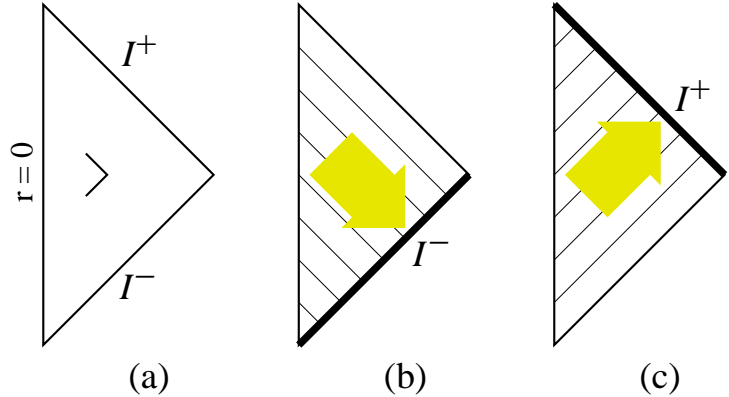}}
\caption%
{\small\sl Like AdS, Minkowski space contains no trapped or
anti-trapped spheres (a).  Unlike AdS, the two allowed null
projections lead to different preferred screens, $\scri^-$ (b) and
$\scri^+$ (c).  Either screen is sufficient to encode the entire
spacetime.  This can be viewed as an expression of the unitarity of
the S-matrix.}
\label{fig-mink}
\end{figure}

For the discussion of scattering processes (Fig.~\ref{fig-mink}) we
shall follow the recipe given in Sec.~\ref{sec-recipe}.
By following past light-cones centered at $r=0$, all of Minkowski
space is projected onto past null infinity, $\scri^-$, where $\theta
\rightarrow 0$.  Similarly, by following future light-cones one can
project the bulk onto future null infinity, $\scri^+$.  Both
infinities are preferred screens of Minkowski space.  Each screen
alone suffices to store all information in the interior of the
space-time; one can interpret this as a statement of the unitarity of
the S-matrix~\cite{Wit98SB} in the absence of black holes.

\begin{figure}[htb!]
  \hspace{.0\textwidth} \vbox{\epsfxsize=1.0\textwidth
  \epsfbox{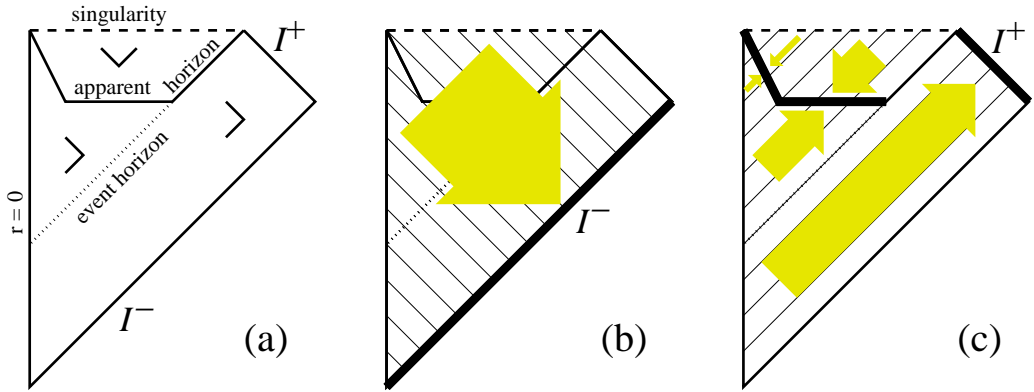}}
\caption%
{\small\sl A classical black hole forms in a scattering process.  The
spheres within the apparent horizon are trapped (a).  All information
can be projected along past light-cones onto $\scri^-$ (b).  But
$\scri^+$ only encodes the information outside the black hole; this
reflects the information loss in the classical black hole.  The black
hole interior can be projected onto the apparent horizon (c).}
\label{fig-mink-bh}
\end{figure}

Let us now assume that a black hole forms during scattering
(Fig.~\ref{fig-mink-bh}a).
The past light-cones still project all points in the spacetime onto
$\scri^-$, including the interior of the black hole
(Fig.~\ref{fig-mink-bh}b).  The screen $\scri^+$, however, encodes
only the exterior of the black hole, via future-directed outgoing
light-rays (Fig.~\ref{fig-mink-bh}c).  This discrepancy can be
interpreted as information loss in classical black holes.  The black
hole interior can be encoded onto the apparent horizon, which forms a
preferred screen, by future-directed outgoing and past-directed
ingoing lightrays.

\begin{figure}[htb!]
  \hspace{.2\textwidth} \vbox{\epsfxsize=.6\textwidth
  \epsfbox{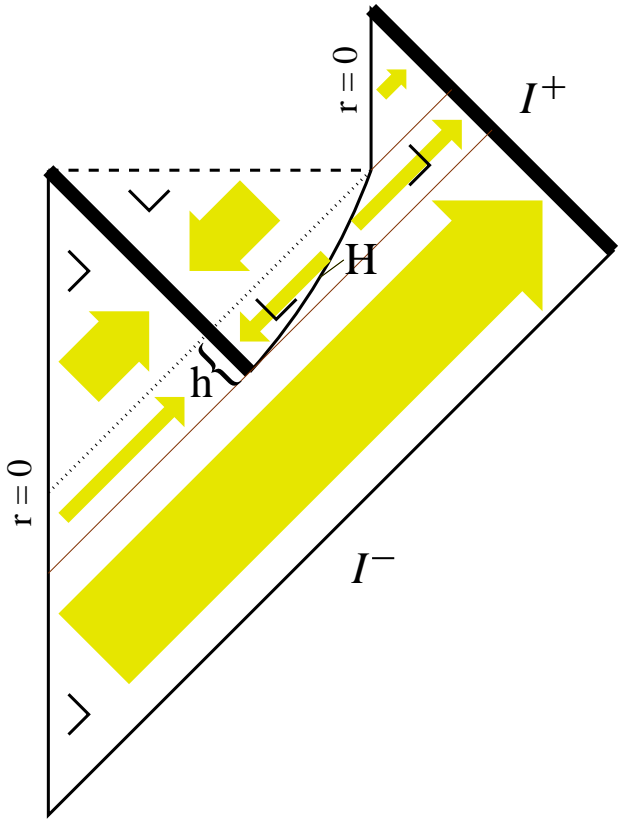}}
\caption%
{\small\sl A quantum black hole forms in a scattering process.
Because negative energy particles cross the apparent horizon during
the evaporating phase ($H$), its size decreases.  The expansion of
future light-cones immediately inside the apparent horizon changes
from negative to positive in this process.  Therefore the maximal area
of the apparent horizon marginally exceeds the maximal area of the
event horizon.  The diagram shows the projection of this space-time
along future light-cones onto screens formed by the apparent horizon
and by $\scri^+$ (thick lines).  The past light-cones would lead to
the usual projection onto $\scri^-$.}
\label{fig-mink-evap}
\end{figure}

The picture becomes more interesting when the quantum radiation of
black holes, as well as its back-reaction, is included.  This restores
the possibility of unitarity.  After the black hole has formed from
classical matter, the apparent horizon shrinks due to the quantum pair
creation of particles~\cite{Haw74}.  In this process a positive energy
particle escapes to infinity, while its negative energy partner
crosses into the black hole.  Unlike positive energy matter, this
particle anti-focusses light: it violates the null convergence
condition and causes the expansion of light-rays to increase.
Outgoing light-rays immediately inside the horizon can thus change
from negative to positive expansion without going through a caustic
(Fig.~\ref{fig-mink-evap}).
(If the null convergence condition~\cite{HawEll} holds, the expansion
becomes positive only at ``caustics,'' or focal points, of the
light-rays.  Caustics thus are the generic endpoints of
light-sheets~\cite{Bou99b}.)  This leads to a situation which would
not be possible in a classical space-time.  There exists a
hypersurface $H$, namely the black hole apparent horizon during
evaporation, from which one has to project {\em away} in {\em both}
directions.  Thus, past-directed ingoing light-rays map $H$ onto a
different part of the apparent horizon ($h$), and future-directed
outgoing light-rays map it onto a part of $\scri^+$.

\subsubsection*{A digression on unitarity}

Let us examine the evaporation process in more detail.  When a
negative mass particle enters the horizon, the expansion of the
generators of the horizon changes from zero to a positive value.
There will be a nearby null congruence, inside the black hole, whose
expansion is changed from a negative value to zero by the same
process.  This congruence will now generate the apparent horizon.
Since it has smaller cross-sectional area, the horizon has shrunk.
The movement of the apparent horizon will leave behind a trace in the
Hawking radiation, causing a deviation from a thermal spectrum over
and above the deviation caused by greybody factors.  This is similar
to the distortion in the thermal spectrum of radiation enclosed in a
cavity, while the wall of the cavity is being moved.

The amount by which the apparent horizon decreases during a given
pair-creation event depends on the {\em profile}, $\theta({\cal A})$,
of the expansion of the outgoing future-directed null geodesics near
the horizon.  The cross-sectional area $\cal A$ of null congruences
becomes smaller, and the expansion $\theta$ more negative, the further
inside the black hole they are located.  If the black hole was formed
by a system of low entropy, for example by the collapse of a
homogeneous dust ball of zero temperature, the profile will be a
featureless monotonic function, and the horizon will decrease very
smoothly during evaporation.  The back-reaction, in this case, will
not imprint a significant signature onto the thermal spectrum.
However, if the black hole was formed by a highly enthropic system,
the profile will be more complex.

Consider a shell of matter falling into a black hole.  For now, assume
that the shell contains only radial modes, and is thus exactly
spherically symmetric even microscopically.  If a lot of entropy is
stored in the shell, its density will be a complicated function of the
radius.  By Raychauduri's equation~\cite{HawEll,Wald}, the density
profile of the infalling shell will be imprinted on the expansion
profile of the outgoing future-directed null geodesics that eventually
pass through the apparent horizon during evaporation.
Correspondingly, the same type of pair creation process will sometimes
cause the horizon area to decrease by a larger step, sometimes by a
smaller amount, depending on the expansion profile of the null
geodesics passing through $H$ at the pair creation event.  The
back-reaction will be irregular, and the corresponding deviations of
the Hawking radiation from the thermal spectrum will be complex.
There is thus a signature in the radiation which encodes the
irregularity of the back-reaction, which in turn encodes the
complexity of the matter system that formed the black hole.

It is easy to extend this discussion to systems containing also
angular modes.  They will deflect outgoing lightrays into angular
directions.  The expansion will now be a local function of the
cross-sectional area, $\theta(\delta {\cal A};\vartheta,\varphi)$.
The back-reaction will not be spherically symmetric, and the apparent
horizon will develop dents and bulges.  This leaves a non-spherical
signature in the Hawking radiation.

In this way information about the material falling into the black hole
may be transferred onto the outgoing Hawking radiation.  The
information will be encoded in a subtle way and it will typically be
necessary to measure the entire radiation emitted by the black hole
before the ingoing state can be reconstructed.  Of course, we have
sketched only a qualitative picture, and we have taken the pair
creation model of black hole evaporation rather literally.  Moreover,
no mechanism can copy ingoing information onto outgoing radiation
unless one implicitly assumes that the fundamental theory evades the
``quantum Xeroxing'' no-go theorem~\cite{SusTho93}, for example by
non-locality~\cite{LowPol95}.%
\footnote{We thank Lenny Susskind for pointing this out, and for
a number of related discussions.}
We have aimed to outline a specific mechanism by which information may
be transferred in the semi-classical picture.  In general terms, our
discussion is strongly related to the approach of
't~Hooft~\cite{DraTho85a,DraTho85b,SteTho94}.

\subsection{de Sitter space}
\label{sec-ds}

de~Sitter space is the maximally symmetric solution of the vacuum
Einstein equation with a positive cosmological constant $\Lambda$.  It
may be visualized as a $(D-1,1)$-hyperboloid embedded in
$(D+1)$-dimensional Minkowski space.  A metric covering the entire
space-time is given by
\begin{equation}
ds^2 = - dt^2 + H^{-2} \cosh^2\! Ht~~ d\Omega_{D-1}^2,
\label{eq-ds-metric}
\end{equation}
where
\begin{equation}
H = \sqrt{\frac{\Lambda}{3}}
\end{equation}
is the Hubble parameter, or inverse curvature radius.  In this metric,
the spacelike hypersurfaces are spheres, $\mathbf{S}^{D-1}$.  They
contract, and then expand, at an exponential rate.  de~Sitter space
also admits metrics with maximally symmetric spatial sections of zero
or negative curvature, as well as a static metric,
\begin{equation}
ds^2 = - (1-H^2 r^2) d\tau^2 + \frac{dr^2}{1-H^2 r^2} + r^2
d\Omega_{D-2}^2.
\end{equation}
Those metrics cover only certain portions of the spacetime.

\begin{figure}[htb!]
  \hspace{-.01\textwidth} \vbox{\epsfxsize=.97\textwidth
  \epsfbox{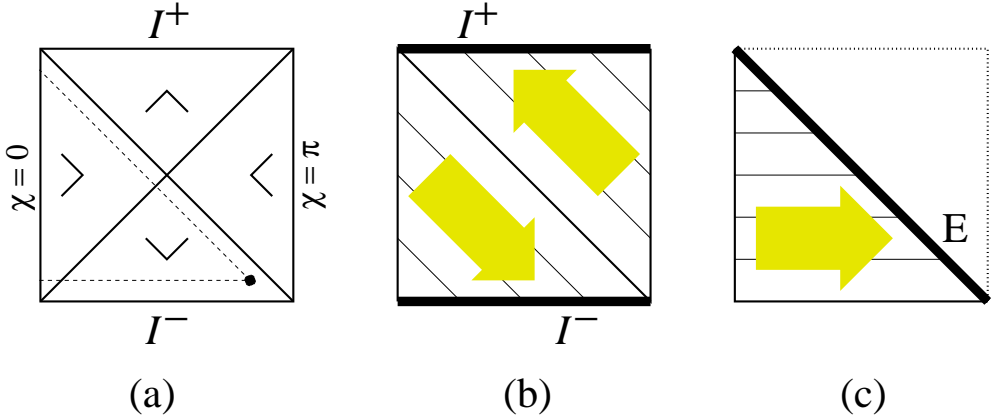}}
\caption%
{\small\sl Penrose diagram for de~Sitter space.  The
$\mathbf{S}^{D-1}$ spacelike slices would correspond to horizontal
lines through the square.  The diagonals are apparent horizons
dividing the space-time into four regions (a).  The $(D-2)$-spheres
near past (future) infinity are trapped (anti-trapped); the spheres
near the poles are normal.  Null projection must be directed towards
the tips of the wedges (see Sec.~\ref{sec-recipe}).  It follows that
de~Sitter space can be projected onto past and future infinity (b),
which are spacelike, optimal screen-hypersurfaces of (exponentially)
infinite size.  A more interesting screen is obtained by applying the
spacelike projection theorem to spheres near the event horizon $E$ of
an observer at $\chi=0$ (a).  By taking a limit, one can show that all
information in the observable region of de~Sitter space can be
projected onto the preferred screen $E$, which is a null hypersurface
of constant spatial area $4\pi H^{-2}$ (c).}
\label{fig-ds}
\end{figure}

The metric on the spatial $(D-1)$-sphere is given by:
\begin{equation}
d\Omega_{D-1}^2 = d\chi^2 + \sin^2\! \chi~d\Omega_{D-2}^2,
\end{equation}
whence
\begin{equation}
r = H^{-1} \cosh Ht~\sin \chi.
\end{equation}
A geodesic observer is immersed in a bath of thermal
radiation~\cite{GibHaw77a} of temperature $T=H/(2\pi)$, which appears
to come from the cosmological horizon surrounding the observer.  The
causal structure of de~Sitter space is shown in Fig.~\ref{fig-ds}.
The only boundaries are past and future infinity, $\scri^-$ and
$\scri^+$; they are both spacelike.  The space-time is divided in half
by the event horizon, $E$, of a geodesic observer, who can be taken to
live at $\chi=r=0$.

Consider the past light-cones centered at $\chi=0$, i.e., at one of
the two poles of the $\mathbf{S}^{D-1}$.  Just as for AdS and
Minkowski, the expansion starts with the value $+\infty$ and
decreases.  Any surface on the light-cone bounds $N_{\rm dof}$ on the
part of the cone it encloses.  By the spacelike projection theorem, it
also bounds $N_{\rm dof}$ on any spatial hypersurface in its interior.
Perhaps surprisingly, this holds even for surfaces which are both near
the event horizon and near the past singularity (see
Fig.~\ref{fig-ds}a, dashed line).  Their area will be $\sim H^{-2}$,
but they enclose an exponentially large spatial region.

When the light-cone reaches $\scri^-$, the expansion approaches zero.
Thus the boundary surface on past infinity is a preferred screen.
(Actually it is optimal because a past light-cone arriving from the
other pole, $\chi=\pi$, will also have $\theta \rightarrow 0$ near
$\scri^-$.)  Repeating this projection for all times, one finds that
half of de~Sitter space, namely the region within the event horizon of
an observer at $\chi=0$, can be projected onto past infinity.  The
projection of only half of the space-time is peculiar to de~Sitter
space.  By contrast, past light-cones project all of Minkowski, or all
of AdS, onto their respective infinities (Secs.~\ref{sec-ads}
and~\ref{sec-mink}).  By using past light-cones emanating from the
other pole of the spatial $\mathbf{S}^{D-1}$, an additional portion of
de~Sitter can be projected onto past infinity.  But this still leaves
out the antitrapped region beyond the cosmological horizons.  It can
only be projected by future-directed lightrays onto future infinity.
A global null projection of de~Sitter space can thus be achieved by
using two optimal screen-hypersurfaces: past and future infinity.
This is shown in Fig.~\ref{fig-ds}b.  Indeed, the potential
holographic role of these boundaries has been speculated upon for some
time~\cite{Wit98SB}.  Both global screens are spacelike hypersurfaces.
Because surfaces near future infinity are anti-trapped, one cannot
encode global de~Sitter space on a finite number of timelike or null
screens.

However, we can apply the spacelike projection theorem
(Sec.~\ref{sec-hp}) to the screens that form the event horizon $E$ of
an observer at $\chi=0$.  They are all of constant area, $4 \pi
H^{-2}$.  (Since $E$ never reaches $\chi=r=0$, this argument strictly
requires a limiting procedure starting from spheres in the vicinity of
$E$; see Fig.~\ref{fig-ds}a.)  Because $E$ is generated by light-rays
of zero expansion, all screens on it are manifestly preferred.  Thus,
the null hypersurface $E$ is a preferred screen of constant area.
Because of its degeneracy, $E$ encodes only itself under null
projection along $E$.  Under spacelike projection, however, it encodes
half of the space-time (Fig.~\ref{fig-ds}c), namely the region within
the event horizon.  One can reasonably argue that the region beyond
the event horizon has no meaning because it cannot be observed, and
that de~Sitter space should not be treated globally~\cite{SusPC,Bou99}
(Fig.~\ref{fig-ds}c); thus the screen $E$ should suffice for a
holographic description of de~Sitter space.

In inflationary models, the de~Sitter phase is followed by a matter or
radiation dominated phase, and the entire space-time during this era
can be projected onto the screens available in the relevant FRW
models~\cite{Bou99b} (see Sec.~\ref{sec-frw}). --- Black holes in
de~Sitter space can be treated much like black holes in AdS or
Minkowski space.

\subsection{FRW cosmologies}
\label{sec-frw}

Friedmann-Robertson-Walker (FRW) cosmologies are described by a metric
of the form
%\begin{equation}
%ds^2 = -dt^2 + a^2(t) \left( \frac{dr^2}{1-kr^2} + r^2 d\Omega^2
%\right).
%\label{eq-FRW1}
%\end{equation}
%Sometimes it is useful to work in comoving coordinates:
\begin{equation}
ds^2 = a^2(\eta) \left[ -d\eta^2 + d\chi^2 + f^2(\chi) d\Omega^2
\right].
\label{eq-FRW2}
\end{equation}
Here
%$k = -1$, $0$, $1$ and
$f(\chi) = \sinh \chi$, $\chi$, $\sin \chi$ corresponds to open, flat,
and closed universes respectively.  FRW universes contain homogeneous,
isotropic spacelike slices of constant (negative, zero, or positive)
curvature.  We will not discuss open universes, since they display no
significant features beyond those arising in the treatment of closed
or flat universes.

The matter content will be described by $T_{ab} =
\mbox{diag}(\rho,p,p,p)$, with pressure $p = \gamma \rho$.  We assume
that $\rho \geq 0$ and $-1/3 < \gamma \leq 1$.  The case $\gamma = -1$
corresponds to de~Sitter space, which was discussed in
Sec.~\ref{sec-ds}.  The {\em apparent horizon} is defined
geometrically as the spheres on which at least one pair of orthogonal
null congruences have zero expansion.  It is given by
\begin{equation}
\eta = q \chi,
\label{eq-ah}
\end{equation}
where
\begin{equation}
q = \frac{2}{1+3 \gamma}.
\end{equation}

\begin{figure}[htb!]
  \hspace{.01\textwidth} \vbox{\epsfxsize=.98\textwidth
  \epsfbox{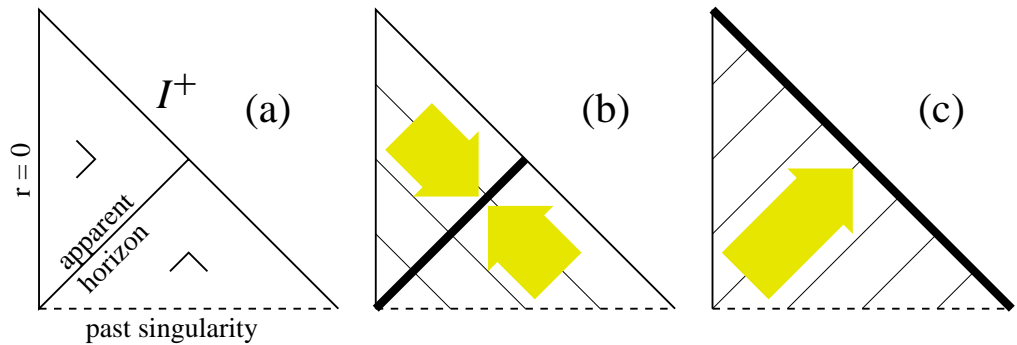}}
\caption%
{\small\sl Penrose diagram for a flat FRW universe dominated by
radiation.  The apparent horizon, $\eta = \chi$, divides the
space-time into a normal and an anti-trapped region (a).  The
information contained in the universe can be projected along past
light-cones onto the apparent horizon (b), or along future light-cones
onto null infinity (c).  Both are preferred screen-hypersurfaces.}
\label{fig-flatfrw}
\end{figure}

The solution for a flat universe is given by
\begin{equation}
a(\eta) = \left( \frac{\eta}{q} \right) ^q.
\end{equation}
Its causal structure is shown in Fig.~\ref{fig-flatfrw}.
The interior of the apparent horizon, $\eta \geq q \chi$, can be
projected along past light-cones centered at $\chi = 0$, or by
space-like projection, onto the apparent horizon.  The exterior, $\eta
\leq q \chi$, can be projected by the same light-cones, but in the
opposite direction, onto the apparent horizon.  The apparent horizon
is thus a preferred screen encoding the entire space-time.
Alternatively, one can use future light-cones to project the entire
universe onto future null infinity, another preferred screen.

By Eq.~(\ref{eq-ah}), the apparent horizon screen is a timelike
hypersurface for $-1/3 < \gamma < 1/3$, null for $\gamma = 1/3$, and
spacelike for $1/3 < \gamma \leq 1$.  In a universe dominated by
different types of matter in different eras, the causal character of
the apparent horizon hypersurface can change from timelike to
spacelike or vice versa (see, e.g., Fig.~5 in Ref.~\cite{Bou99b}).

\begin{figure}[htb!]
  \hspace{.1\textwidth} \vbox{\epsfxsize=.8\textwidth
  \epsfbox{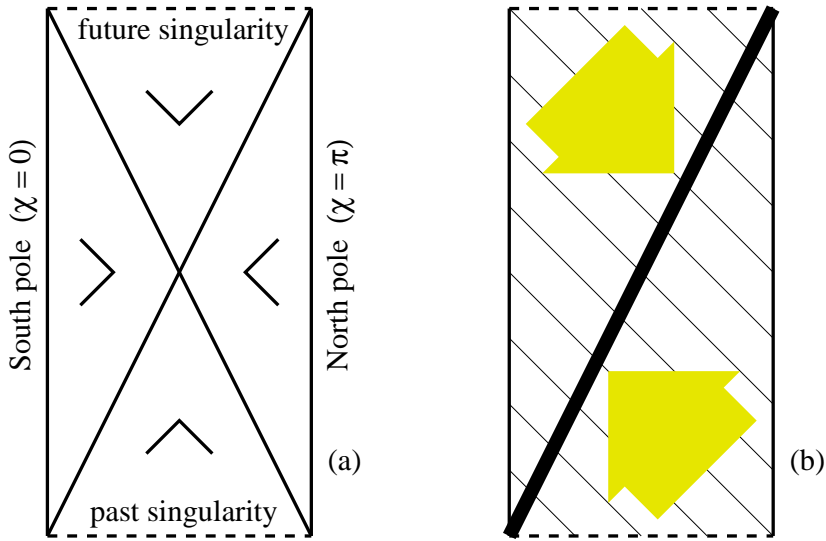}}
\caption%
{\small\sl Penrose diagram for a closed FRW universe dominated by
pressureless dust. Two apparent horizons divide the space-time into
four regions (a).  The information in the universe can be projected
onto the embedded screen-hypersurface formed by either horizon (b).}
\label{fig-closedfrw}
\end{figure}

For a closed universe, the solution is given by
\begin{equation}
a(\eta) = a_{\rm max} \left( \sin \frac{\eta}{q} \right)^q.
\end{equation}
In addition to Eq.~(\ref{eq-ah}), a second apparent horizon
emanates from the opposite pole of the spatial $\mathbf{S}^{D-1}$, at
$\chi=\pi$; it is described by
\begin{equation}
\eta = q (\pi-\chi).
\label{eq-ah2}
\end{equation}
The two hypersurfaces formed by the apparent horizons divide the
space-time into four regions, as shown in Fig.~\ref{fig-closedfrw}.
Let us choose the first apparent horizon, Eq.~(\ref{eq-ah}), as a
(preferred) screen-hypersurface.  On one side, $\eta \geq q \chi$,
lies a normal region and a trapped region.  These regions can be
projected onto the screen by past-directed radial light-rays moving
away from the South pole ($\chi=0$).  The other half of the universe,
$\eta \leq q \chi$, can be projected onto the same screen by future
directed radial light-rays moving away from the North pole
($\chi=\pi$).  Therefore the preferred screen given by
Eq.~(\ref{eq-ah}) encodes the entire closed universe.

A number of cosmological entropy bounds have been
proposed~\cite{EasLow99,Ven99,BakRey99,KalLin99,Bru99} which are
based on the idea of defining a horizon-size spatial region to which
Bekenstein's bound can be directly applied.  We have emphasized the
importance of these bounds in Ref.~\cite{Bou99b}, where we also
discuss their relation to the covariant entropy bound
(Sec.~\ref{sec-hp}).  Those bounds can be given a holographic
interpretation by considering them as limits on $N_{\rm dof}$ in the
specified ken.  Because they refer to limited regions, however, it is
not clear how global screen-hypersurfaces could be constructed.

\subsection{Einstein static universe}
\label{sec-esu}

\begin{figure}[htb!]
  \hspace{.15\textwidth} \vbox{\epsfxsize=.7\textwidth
  \epsfbox{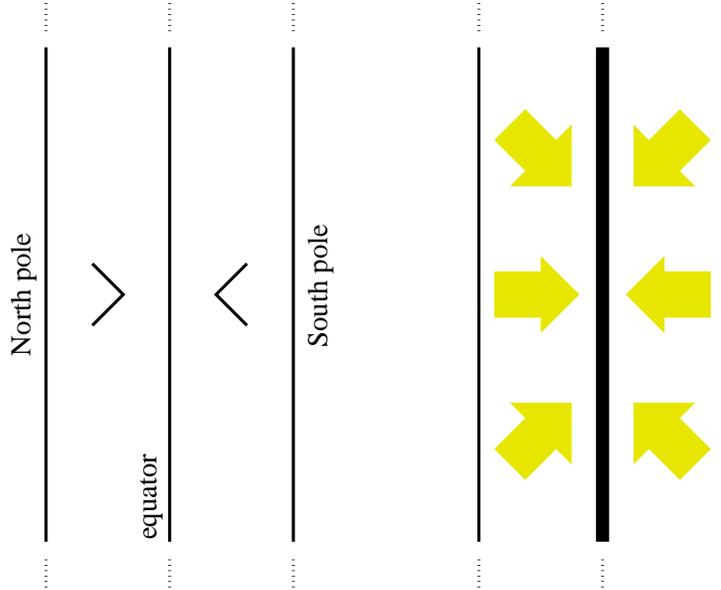}}
\caption%
{\small\sl Penrose diagram for the Einstein static universe.  The
equator separates two normal regions.  It forms an optimal, timelike
screen-hypersurface of constant area, encoding all information by null
or spacelike projection.  These properties are shared by the boundary
of AdS.}
\label{fig-esu}
\end{figure}

The Einstein static universe (ESU) is a closed FRW space-time
containing ordinary matter as well as a positive cosmological constant
of a certain critical value~\cite{HawEll,MTW}.  Its metric can be
written as a direct product of an infinite time axis with a
$(D-1)$-sphere of constant radius $a$:
\begin{equation}
ds^2 = - dt^2 + a^2 d\Omega_{D-1}^2.
\end{equation}
The causal structure is shown in Fig.~\ref{fig-esu}.
Each hemisphere can be projected along past- or future-directed
light-rays, or by spacelike projection, onto the equator.  This screen
is optimal, because all four families of orthogonal light-rays have
vanishing expansion.  Moreover, the screen forms a timelike
hypersurface, with spatial slices of constant finite size.

This is reminiscent of the properties of the screen at the boundary of
Anti-de~Sitter space: the screen is optimal, timelike, of constant
size, and encodes the entire space-time by space-like projection.  The
difference is that the AdS screen has infinite proper area, while the
equator in the ESU is a $(D-2)$-sphere of finite area $\sim a^{D-2}$.
It lies not on a boundary of space (there is none in the ESU), but is
embedded in the interior.

The properties of the projection might give rise to the hope that a
boundary theory, dual to the bulk description, could be formulated on
the equator of the ESU.  The example may be of limited use, however,
because the ESU is not a stable solution~\cite{HawEll,MTW}.  Another
unstable solution with similar properties is given by a static
spherical system just on the verge of gravitational collapse.  Its
radius will be equal to its gravitational radius, and the expansion of
both past- and future-directed outgoing light-rays goes to zero at the
surface of the system.

\pagebreak

%%%%%%%%%%%%%%%%%%%%%%%%%%%%%%%%%%%%%%%%%%%%%%%%%%%%%%%%%%%%%%%%%%%%
\sect{Holographic Theory}
\label{sec-discussion}
%%%%%%%%%%%%%%%%%%%%%%%%%%%%%%%%%%%%%%%%%%%%%%%%%%%%%%%%%%%%%%%%%%%%

\subsection{Summary}
\label{sec-summary}

From a universal entropy bound found in Ref.~\cite{Bou99b}, we
obtained a background-independent formulation of the holographic
principle~\cite{Tho93,Sus95}.  This led us to a construction of
hypersurfaces (screens) on which all information contained in a
space-time can be stored.  The screens are embedded, or lie on the
boundary of the space-time, and contain no more than one bit of
information per Planck area.  In this sense, the world is a hologram.

The construction was applied to a number of examples.  For
Anti-de~Sitter space it yields the timelike boundary at spatial
infinity as a preferred screen.  In Minkowski space, past or future
null infinity, or a flat plane, can encode all information.  de~Sitter
space is mapped along light-rays onto the spacelike infinities in the
past and future; alternatively, all information in the observable half
of the de~Sitter space can be stored on the event horizon (a null
hypersurface of finite area) via spacelike projection.  Cosmological
spacetimes may not have a boundary, but embedded screens can be found;
they may be spacelike, timelike, or null, depending on the matter
content.  The information in a black hole can be mapped onto the
apparent horizon, or onto past null infinity.

From the examples one can draw the following observations:
\begin{itemize}

\item{Holographic screens can be spacelike hypersurfaces.}

\item{If they are timelike or null, the spatial area is not
necessarily constant in the induced, or any other, time-slicing.}

\end{itemize}
Before explaining why these features may be significant, let us
briefly discuss a tempting but misguided conclusion.  One might argue
that holographic projection onto spacelike screens is a trivial
accomplishment, because in any conventional theory one can specify
initial conditions on a Cauchy surface and predict, or retrodict, the
past and future development of the system.  That is true, but it is a
different kind of information storage.  In that case, one stores not
only the information at one moment of time, but also a machine (namely
the theory) which is capable of recovering the state of the system at
all other times.  A holographic construction, on the other hand,
feigns ignorance of any theory describing the matter evolution, and
simply encodes all information, at all times, onto screens of
dimension $D-2$.  The space-time is sliced into null hypersurfaces;
slice by slice the information is encoded onto $(D-2)$-dimensional
spatial surfaces at a maximum density of one bit per Planck area.
These surfaces form a $(D-1)$ dimensional screen hypersurface which
may be timelike, spacelike, or null, but from the point of view of
holographic information storage its causal character is irrelevant.

\subsection{Theories on the screen}
\label{sec-screentheories}

Our interpretation, so far, has centered on the {\em information}
needed to describe a state.  This is measured by the number of degrees
of freedom.  Therefore we can use the holographic principle (which
refers to $N_{\rm dof}$) to project all information in the space-time
onto screen-hypersurfaces.  In this sense, the holographic principle
implies a drastic reduction of the complexity of nature compared to
naive expectations of, perhaps, one degree of freedom per Planck
volume.

We did not, however, use the holographic principle to {\em describe}
nature.  The holographic principle is far from manifest in the
description of the world in terms of general relativity and quantum
field theory; yet these theories are very successful.  Working within
their frame, one finds a number of non-trivial effects which appear to
insure that the entropy bound implied by the holographic principle is
always satisfied~\cite{Bou99b}; but these results could not have been
immediately inferred from the basic axioms of GR and QFT.

As a kind of external restriction imposed on physical theories,
holography is interesting but unsatisfactory.  If the number of
degrees of freedom is limited by the holographic principle, there
ought to be a description of nature in which this restriction is
manifest.  Let us call this hypothetical description {\em the
holographic theory}.  One would expect the holographic theory to
remain valid when semi-classical gravity breaks
down~\cite{Tho93,Mal97}; in this regime it may be the only possible
description.  These are good reasons to search for a holographic
theory.

The simplest idea would be to define a theory on the geometric
background given by the screen-hypersurface(s).  If the theory
contained one degree of freedom per Planck area, and was related by a
kind of dictionary (``duality'') to the space-time (``bulk'') physics,
the holographic principle would be manifest.  Let us call this type of
theory a {\em dual theory}.

This idea works for certain asymptotically Anti-de~Sitter space-times.
The screen encoding the entire bulk information is the timelike
hypersurface formed by the boundary of space (Sec.~\ref{sec-ads}).
According to Maldacena's remarkable conjecture~\cite{Mal97}, a
super-Yang-Mills theory living on this hypersurface describes the bulk
physics completely.  By considering a finite boundary and taking the
limit as it moves to spatial infinity, one can show that the theory
contains no more than one degree of freedom per Planck
area~\cite{SusWit98}.  Therefore it is a dual theory in the sense of
our definition.

Perhaps it will be possible to find dual theories for some other
classes of space-times; certainly this would be be an important
contribution to the understanding of holography and of quantum
gravity.  In general, however, the ``dual theory'' approach will not
work.  The theories we usually think of have a fixed number of degrees
of freedom built into them; these degrees of freedom evolve in
Lorentzian time.  But consider the cosmological solutions studied in
Sec.~\ref{sec-frw}.  The area of the screens is time-dependent.  The
screen theory would have to be capable of ``creating'' or
``activating'' degrees of freedom.  Moreover, the area can decrease,
as seen in the closed universe example.  In the screen theory this
would correspond to the destruction, or de-activation, of degrees of
freedom.  Eventually their number would approach zero, and the second
law of thermodynamics would be violated in the screen theory.%
\footnote{We are grateful to Andrei Linde for stressing this point to
us.}
(Note that this does not, of course, imply a violation of the second
law in the bulk.  Rather, it is related to the creation, or
destruction, of degrees of freedom at the initial and final
singularities of the universe.)

This suggests that one should not in general think of the screen
theory as a conventional theory with a fixed number of degrees of
freedom.  Rather, one might expect it to be a theory with a varying
number of ``active'' degrees of freedom.  Thus, its properties would
be very different from those of ordinary physical theories.  Moreover,
since the screen hypersurfaces can be spacelike or null, one should
not expect the theory to live in Lorentzian time.

\subsection{Geometry from entropy}
\label{sec-ge}

We would like to advocate a more radical approach.  One should not be
thinking about a ``screen theory'' (a theory defined on some
hypersurface of space-time) at all.  The screen theory approach cannot
be fundamental, because it presumes the existence of a space-time
background, or at least of an asymptotic structure of space and time.
In order to use the holographic principle for a full description of
nature, we suggest it should be turned around.  Loosely speaking, one
should not constrain entropy by geometry, but construct geometry from
entropy.  (Strictly, ``number of degrees of freedom'' should replace
``entropy'' here.)  The construction must be such that the holographic
principle, in the form given in Sec.~\ref{sec-hp}, is automatically
satisfied.

It thus appears that two problems must be overcome if a holographic
theory is to be found.  First, one must formulate a theory with a
varying number of degrees of freedom.  A possibility may be that the
theory can activate or de-activate degrees of freedom from an infinite
reservoir.%
\footnote{We thank Lenny Susskind for this suggestion and related
discussions.}
An extreme but perhaps more satisfying resolution would be to treat
quantum degrees of freedom not as fundamental ingredients, but as a
derived concept.  't~Hooft has long been advocating that models should
be sought in which quantum degrees of freedom arise as a complex,
effective structures (see Ref.~\cite{Tho99} and references therein).
It would be natural for $N_{\rm dof}$ to vary in such models.

The second challenge is to find a prescription that allows the unique
reconstruction of space-time geometry from the varying number of
degrees of freedom (see, e.g., Refs.~\cite{Jac94,MarSmo98} and
references therein for a discussion of related questions).  Part of
this prescription will be to equate $N_{\rm dof}$ with the proper area
of an embedded $(D-2)$-dimensional preferred or optimal screen.  A
more difficult question is how the intrinsic geometry of
screen-hypersurfaces can be recovered.  It may be undesirable to
identify the discrete steps of, say, a cellular
automaton~\cite{Tho88,Tho90} with Lorentzian time.  But the number and
character of the degrees of freedom provide information about the
matter content.  Therefore a complete reconstruction of space-time
geometry is not inconceivable.

\section*{Acknowledgments}

I am indebted to Gerard 't~Hooft, Andrei Linde, and Lenny Susskind for
very helpful discussions, as well as important criticism and
suggestions.  I thank Lenny Susskind for valuable comments on a draft
of this paper.

%\bibliographystyle{board}
%\bibliography{all}

\end{document}